# REVIEW AND ANALYSIS OF LOCAL MULTIPOINT DISTRIBUTION SYSTEM (LMDS) TO DELIVER VOICE, DATA, INTERNET, AND VIDEO SERVICES

Dr.S.S.Riaz Ahamed.
Principal, Sathak Institute of Technology, Ramanathapuram,TamilNadu, India-623501.
Email:ssriaz@ieee.org

**ABSTRACT**

Local multipoint distribution system (LMDS) uses cellular-like network architecture of microwave radios placed at the client's location and at the company's base station to deliver fixed services, mainly telephony, video and Internet access. The use of time-division multiple access (TDMA) and FDMA (frequency DMA) technology allows multiple customers within a 3-5 mile coverage radius to share the same radio channel. Customers can receive data rates between 64kbps to 155Mbps. LMDS was conceived as a broadband, fixed wireless, point-to-multipoint technology for utilization in the last mile. Throughput capacity and reliable distance of the link depends on common radio link constraints and the modulation method used - either phase-shift keying or amplitude modulation. In general deployment links of up to 5 miles (8 km) from the base station are possible, but distance is typically limited to about 1.5 miles due to rain fading attenuation constraints. Point-to-point systems are also capable of using the LMDS frequencies and can reach slightly farther distances due to increased antenna gain.LMDS uses a scalable architecture combined with industry standards to ensure service can be expanded as customer demand increases.

## 1. INTRODUCTION

Local multipoint distribution system (LMDS) is the broadband wireless technology used to deliver voice, data, Internet, and video services in the 25-GHz and higher spectrum (depending on licensing). As a result of the propagation characteristics of signals in this frequency range, LMDS systems use a cellular-like network architecture, though services provided are fixed, not mobile. In the United States, 1.3 MHz of bandwidth (27.5 B 28.35 GHz, 29.1 B 29.25 GHz, 31.075 B 31.225 GHz, 31 B 31.075 GHz, and 31.225 B 31.3 GHz) has been allocated for LMDS to deliver broadband services in a point-to-point or point-to-multipoint configuration to residential and commercial customers[1]-[4][21]-[27].

LMDS is a broadband wireless access technology governed by the IEEE and is outlined by the 802 LAN/MAN Standards Committee through the efforts of the IEEE 802.16.1 Task Group. LMDS commonly operates on microwave frequencies across the 26GHz and 29GHz bands. In the United States, frequencies from 31.0 through 31.3 GHz are also considered LMDS frequencies. The acronym LMDS is derived from the following:

- L (Local) denotes that propagation characteristics of signals in this frequency range that limit the coverage area of a single hub site to a smaller footprint than that of other broadcast transmitters. Several such hubs can be created to provide the desired coverage area in a metropolitan area;
- M (Multipoint) indicates that the broadcast signals are available to multiple users, ie, signal is shared by multiple subscribers; the wireless return path, from subscriber to the base station, is a direct point-to-point transmission;
- D (Distribution) refers to the distribution of signals, which may consist of simultaneous voice, data, Internet, and video traffic to targeted subscribers;
- S (System) implies the system provides the operator a choice to serve any subscriber with the desired services that will be profitable today with adequate flexibility to grow and change based on evolving demands.

## 2. ADVANTAGE OF THE LMDS TECHNOLOGY

The biggest advantage of the LMDS technology is derived from its wireless nature: it gives operators an opportunity to provide consumers with access bandwidths and two-way capability without digging up streets. This one fact that no right-of-way clearances are required to deploy it makes it sufficiently enticing as a technology. LMDS has huge chunks of under-utilised spectrum ready to go, more than enough to satisfy pent-up demand for voice, private data networks, and high-speed Internet access. Further, the technology is stable and all-digital, and assures effective delivery of high-quality services that are demanded by consumer and enterprise users.

In addition, this advantage has provided today cellular operators across the globe an economical way to quickly deploy cell sites to provide next generation of bandwidth intensive data services, beyond the traditional voice service. This added benefit will certainly not be lost in India’s competitive landscape with cellular and limited-mobility operators going head-to-head and where the fine line between fixed and cellular services blends into converged applications for the benefit of the consumer[2][4][7]-[11].

## 3. CONCEPT

LMDS is a broadband wireless point-to-multipoint communication system operating above 20 GHz (depending on country of licensing) that can be used to provide digital two-way voice, data, Internet, and video services (see *Figure*).

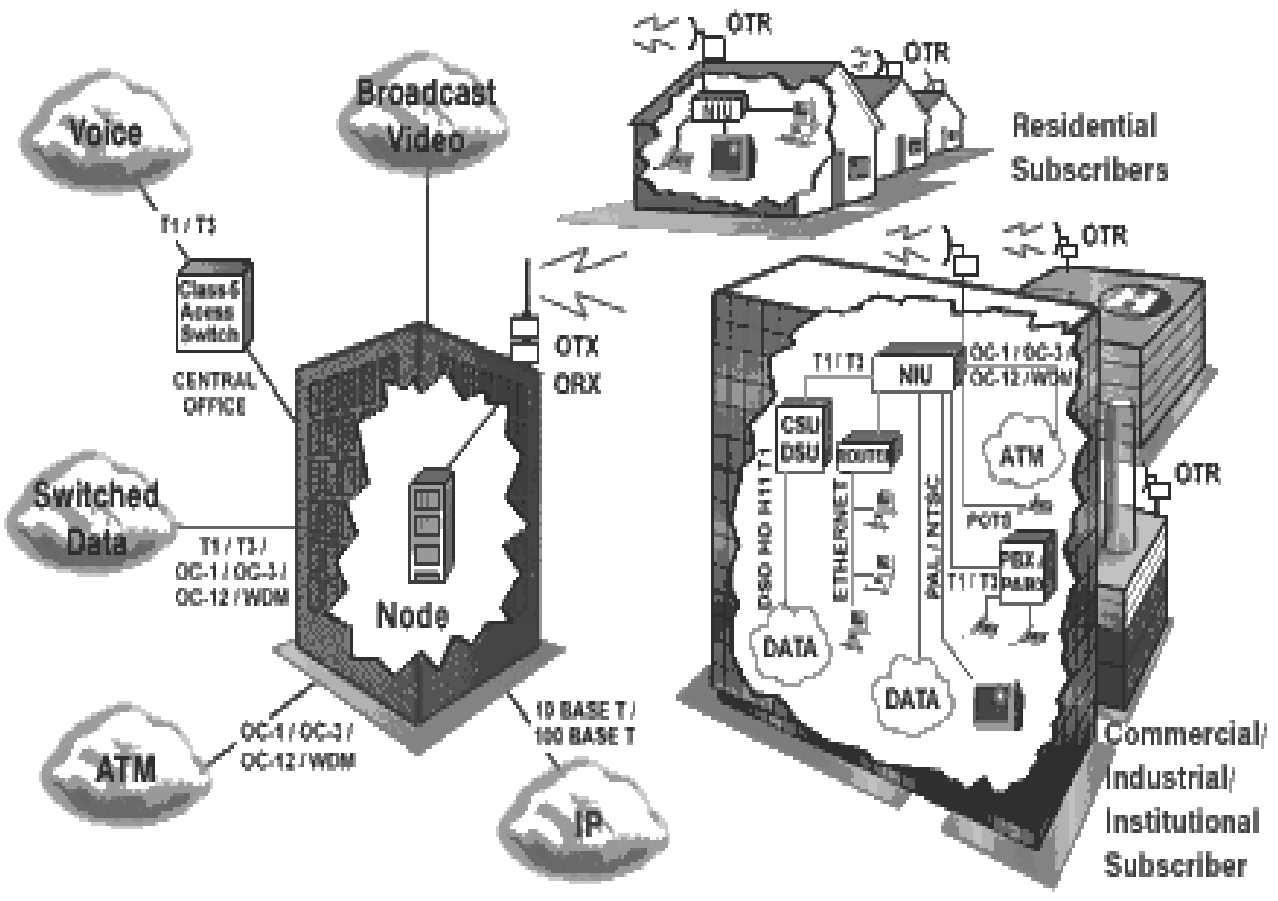

**Figure1. LMDS System**

## 4. WHY LMDS?

Point-to-point fixed wireless networks have been commonly deployed to offer high-speed dedicated links between high-density nodes in a network. More recent advances in a point-to-multipoint technology offer service providers a method of providing high-capacity local access that is less capital-intensive than a wireline solution, faster to deploy than wireline, and able to offer a combination of applications. Moreover, as a large part of a wireless network's cost is not incurred until the customer premises equipment (CPE) is installed, the network service operator can time capital expenditures to coincide with the signing of new customers[2]-[9]. LMDS provides an effective last-mile solution for the incumbent service provider and can be used by competitive service providers to deliver services directly to end users. Benefits can be summarized as follows:

- lower entry and deployment costs
- ease and speed of deployment (systems can be deployed rapidly with minimal disruption to the community and the environment)
- fast realization of revenue (as a result of rapid deployment)
- demand-based buildout (scalable architecture employing open industry standards ensuring services and coverage areas can be easily expanded as customer demand warrants)
- cost shift from fixed to variable components (with traditional wireline systems, most of the capital investment is in the infrastructure, while with LMDS a greater percentage of the investment is shifted to CPE, which means an operator spends dollars only when a revenue-paying customer signs on)

- no stranded capital when customers churn
- cost-effective network maintenance, management, and operating costs

## 5. NETWORK ARCHITECTURE

Various network architectures are possible within LMDS system design. The majority of system operators will be using point-to-multipoint wireless access designs, although point-to-point systems and TV distribution systems can be provided within the LMDS system. It is expected that the LMDS services will be a combination of voice, video, and data. Therefore, both asynchronous transfer mode (ATM) and Internet protocol (IP) transport methodologies are practical when viewed within the larger telecommunications infrastructure system of a nation. The LMDS network architecture consists of primarily four parts: network operations center (NOC), fiber-based infrastructure, base station, and CPE[14]-[20].

## 6. SYSTEM EQUIPMENT SEGMENTS

The NOC contains the network management system (NMS) equipment that manages large regions of the customer network. Multiple NOCs can be interconnected. The fiber-based infrastructure typically consists of synchronous optical network (SONET) optical carrier (OC)–12, OC–3, and DS–3 links; central-office (CO) equipment; ATM and IP switching systems; and interconnections with the Internet and public switched telephone networks (PSTNs).

The base station is where the conversion from fibered infrastructure to wireless infrastructure occurs. Base-station equipment includes the network interface for fiber termination; modulation and demodulation functions; and microwave transmission and reception equipment typically located atop a roof or a pole. Key functionalities which may not be present in different designs include local switching. If local switching is present, customers connected to the base station can communicate with one another without entering the fiber infrastructure. This function implies that billing, channel access management, registration, and authentication occur locally within the base station.

The alternative base-station architecture simply provides connection to the fiber infrastructure. This forces all traffic to terminate in ATM switches or CO equipment somewhere in the fiber infrastructure. In this scenario, if two customers connected to the same base station wish to communicate with each other, they do so at a centralized location. Billing, authentication, registration, and traffic-management functions are performed centrally.

The customer-premises configurations vary widely from vendor to vendor. Primarily, all configurations will include outdoor-mounted microwave equipment and indoor digital equipment providing modulation, demodulation, control, and customer-premises interface functionality. The CPE may attach to the network using time-division multiple access (TDMA), frequency-division multiple access (FDMA), or code-division multiple access (CDMA) methodologies. The customer premises interfaces will run the full range from digital signal, level 0 (DS–0), plain old telephone service (POTS), 10BaseT, unstructured DS–1, structured DS–1, frame relay, ATM25, serial ATM over T1, DS–3, OC–3, and OC–1. The customer premises locations will range from large enterprises (e.g., office buildings, hospitals, campuses), in which the microwave equipment is shared between many users, to mall locations and residences, in which single offices requiring 10BaseT and/or two POTS lines will be connected. Obviously, different customer-premises locations require different equipment configurations and different price points [3][4][9][11]-[17].

## 7. ARCHITECTURAL OPTIONS

LMDS system operators offer different services and have different legacy systems, financial partners, and business strategies. As a result, the system architecture used will differ between all system operators. The most common architectural type uses co-sited, base-station equipment. The indoor digital equipment connects to the network infrastructure, and the outdoor microwave equipment mounted on the rooftop is housed at the same location (see Figure). Typically, the radio frequency (RF) planning for these networks uses multiple sector microwave systems, in which transmit- and receive-sector antennas provide service over a 90-, 45-, 30-, 22.5-, or 15-degree beamwidth. The idealized circular coverage area around the cell site is divided into 4, 8, 12, 16, or 24 sectors.

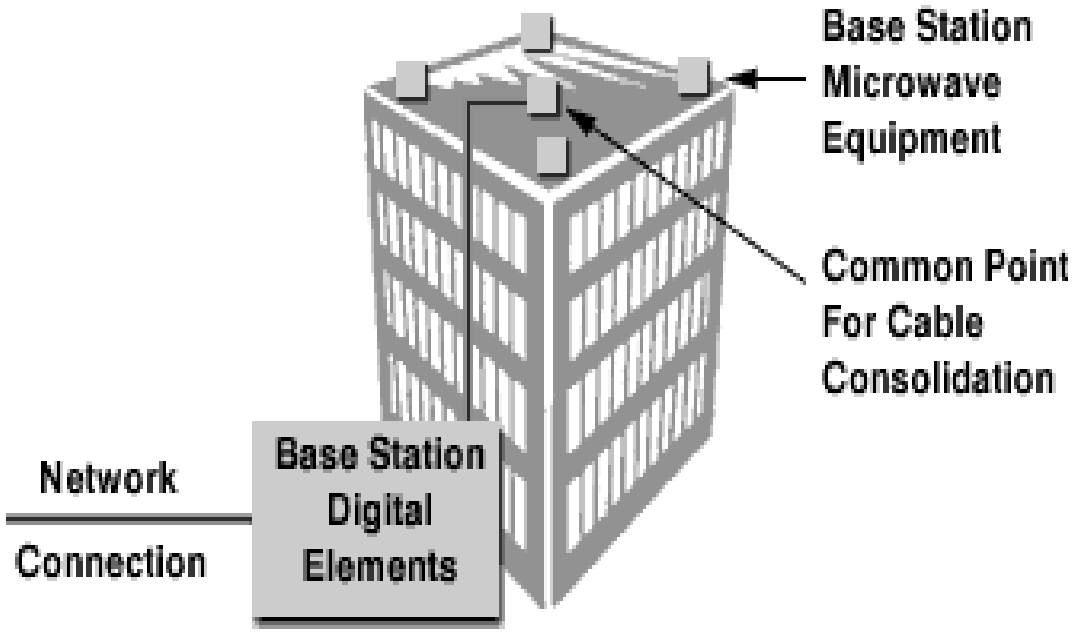

***Figure2. Co-Sited Base Station***

Alternative architectures include connecting the base-station indoor unit to multiple remote microwave transmission and reception systems with analog fiber interconnection between the indoor data unit (IDU) and outdoor data unit (ODU). This approach consolidates the digital equipment, providing increased redundancy, reduced servicing costs, and increased sharing of digital resources over a larger area. The difficulties are typically the lack of analog fiber resources and remote microwave transmission and reception equipment deployment issues. By using remote microwave equipment, there may be a reduced sectorization requirement at each remote location. This second alternative architecture is early in the design process for most vendors and standards bodies (see Figure) [2]-[7].

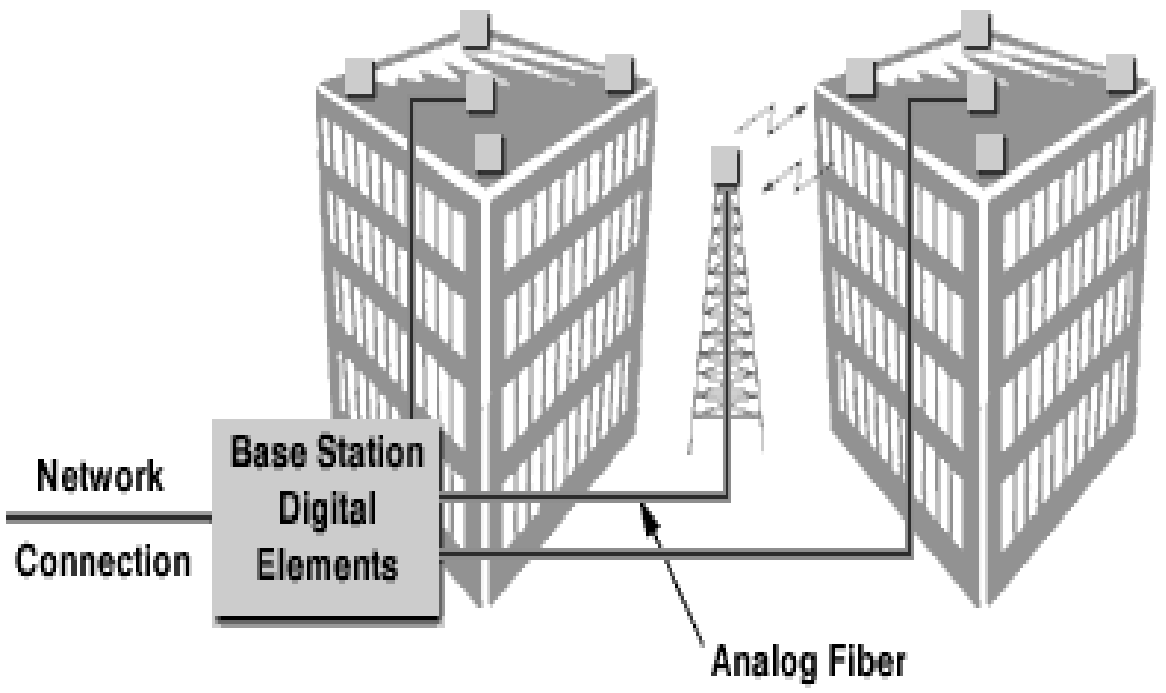

***Figure3. Analog Fiber Architecture***

## 8. MODULATION

Modulation methods for broadband wireless LMDS systems are generally separated into phase shift keying (PSK) and amplitude modulation (AM) approaches. The modulation options for TDMA and FDMA access methods are almost the same.

The TDMA link modulation methods typically do not include the 64–quadrature amplitude modulation (QAM), although this might become available in the future. The FDMA access modulation methods are listed in Table 2 and are rated on an estimated scale as to the amount of bandwidth they require for a 2–Mbps constant bit rate (CBR) connection (without accounting for overhead due to ATM and FEC). Values are approximate, as there are issues involving channel filter mask roll-off factors, which can be important when providing the relationship between microwave bandwidth and data rates[10]-[19].

| Name | Modulation Method | MHz for 2 Mbps CBR Connection |
|---|---|---|
| BPSK | binary phase shift keying | 2.8 MHz |
| DQPSK | differential QPSK | 1.4 MHz |
| QPSK | quaternary phase shift keying | 1.4 MHz |
| 8PSK | octal phase shift keying | 0.8 MHz |
| 4–QAM | quadrature amplitude modulation, 4 states | 1.4 MHz |
| 16–QAM | quadrature amplitude modulation, 16 states | 0.6 MHz |
| 64–QAM | quadrature amplitude modulation, 64 states | 0.4 MHz |

**Table1. FDMA Access Modulation Methods**

## 9. NETWORK MANAGEMENT

LMDS network management is designed to meet a network operator's business objectives by providing highly reliable network management services. Network management requires the following:

**Fault Management**

This is necessary to identify, localize, and correct errors or faults in the network. Each device within a wireless network should be monitored for troubleshooting or performance. All LMDS devices collect and report statistics pertaining to traffic throughput, boundary condition violations, and management activities.

**Configuration Management**

This is necessary in order to provision, inventory, initialize, and back-up network resources. The LMDS equipment should be auto-discovered when new equipment is added to a node. This minimizes the amount of provisioning needed to install or upgrade equipment.

**Accounting Management**

This is necessary to collect and process billing information. Each manageable node in the wireless portion of the network should maintain a collection of statistics that can be accessed by a third-party billing system as input. Users should be identified on a per-network user basis.

**Performance Management**

This is necessary to collect, filter, and analyze network resource statistics. There are a number of parameters that should be monitored and configured on each network node, from T1 traffic throughput to output power level. The management station should monitor these parameters and adjust them to increase performance.

**Security Management**

All information transmitted through the wireless environment must be encrypted between each node in the network. The security-management function should automatically generate and coordinate the keys used to encrypt and decrypt, as well as to authenticate users.

The management application at the base station should not be a stand-alone management application. It must provide a mechanism to populate the cell-based information in the node's management information base.

For the next few years, dedicated platforms may be required to provide end-to-end management of the complete LMDS system [7] [16] [19] [21]-[23].

## 10. CONCLUSION

Via the transmission of microwave signals, LMDS networks can provide two-way broadband services including video, high-speed Internet access, and telephony services. LMDS leap frogs all the hurdles encountered when wired networks must be deployed. It really could be the answer to the long sought solution to last mile broadband in the world. Breakthroughs in radio technology, along with increased industry confidence following the success of personal communications service (PCS) and cellular mobile services, have dramatically improved confidence in radio as a reliable local access technology. LMDS, as a technology is mature and reliable and delivers compelling advantages for adoption in the world.